\documentclass[10pt,a4paper]{llncs}

\usepackage[T1]{fontenc}
\usepackage[utf8]{inputenc}
\usepackage{tgheros}
\usepackage[a4paper, lmargin=2cm, rmargin=2cm, tmargin=2.8cm, bmargin=2.8cm]{geometry}

\usepackage{mathtools}
\mathtoolsset{showonlyrefs=true}

\usepackage{hyperref}
\usepackage{amsmath}
\usepackage{amssymb}
\usepackage{color}
\usepackage{graphicx}
\usepackage{enumerate}
\usepackage{authblk}
\usepackage{caption}
\usepackage{subcaption}

\DeclareMathOperator*{\argmax}{arg\,max}

\spnewtheorem{assumption}{Assumption}{\bfseries}{}

\title{Inefficiency of CFMs: hedging perspective and agent-based simulations}

\author{Anonymous for submission}
\author{Samuel Cohen$^{1,3}$, Marc Sabat\'e Vidales$^{2,3}$, David Siska$^{2,4}$ and {\L}ukasz Szpruch$^{2,3}$}
\institute{$^{1}$Mathematical Institute, University of Oxford\\ 
$^{2}$School of Mathematics, University of Edinburgh\\
$^{3}$Simtopia.ai\\
$^{4}$Vega Protocol
}
\date{\today}

\begin{document}
\maketitle 



\begin{abstract}
We investigate whether the fee income from trades on the
CFM is sufficient for the liquidity providers to hedge away the exposure to market risk. 
We first analyse this problem through the lens of continuous-time financial mathematics and derive an upper bound for not-arbitrage fee income that would make CFM efficient and liquidity provision fair. We then evaluate our findings by performing multi-agent simulations by varying CFM fees, market volatility, and rate of arrival of liquidity takers. 
We observe that, on average, fee income generated from liquidity provision is insufficient to compensate for market risk. 
\end{abstract}

\section{Introduction}

It has been shown empirically~\cite{faycal} and experimentally~\cite{sabate2022variable} that liquidity providers (LPs) in constant function markets (CFMs) (in particular, Uniswap v3) lose money on average. Indeed, from e.g.~\cite[Table 2]{faycal}, the average LP transaction in the ETH/USDC pool (from May 2021–-August 2022) resulted in a position loss of $-1.64\%$, and fee income of $0.155\%$ of
the size of the initial trade, with an average hold time of 6.1 days.

An LP entering a position in CFM opens themselves to two risks: the impermanent loss due to price moves in the underlying asset and the risk of fee income not compensating the impermanent loss. 
A rational LP may choose to hedge the impermanent loss component, which is essentially a perpetual option~\cite{angeris2022primer}.
If the fee income (from trades on the CFM) exceeds the cost of the hedge the LP is making a risk-free profit.
However, if the fee income is below the cost of the hedge the LP is making a loss. 
We formalise this intuition using tools from continuous-time financial mathematics, to derive a theoretical, arbitrage-free upper bound on the fee income that would make the CFM efficient and LP positions fair. 
This is our Theorem~\ref{thm critical rate}.
This is complementary perspective to that of loss versus rebalancing (LVR) introduced in~\cite{milionis2022automated}.

The fee income of the LP depends on the CFM fee (denoted $1-\gamma$ in Uniswap) and the trade flow.
There is no canonical way to jointly model the underlying price process and the trade flow which realistically captures the interdependence of the two.  For this reason we analyse the relationship between the LP fee income and the arbitrage-free upper bound using simulations involving a CFM, another liquid exchange, an arbitrageur agent and price sensitive noise traders (they choose between the exchanges based on which offers better execution price). 
We considered the CFM fee to be in the range $0\%$ to $4\%$ and in this range the simulations show that fee income due to the LP never exceeds the cost of hedging i.e. the LP is always losing money.
As one would expect from Theorem~\ref{thm critical rate} the loss is higher if the volatility of the underlying asset is higher. 
This is regardless of the rate of noise traders arriving in the market. 
One may hope that the solution is to increase the fee the CFM charges but this reduces the trade flow thus reducing the income. 
Similarly, reducing the slippage by depositing more into the pool would increase trade but would not lead to lower losses for LPs as this increases the hedging cost of the LP position.

\begin{figure}[h!]
     \centering
     \begin{subfigure}[b]{0.45\textwidth}
         \centering
         \includegraphics[width=\textwidth]{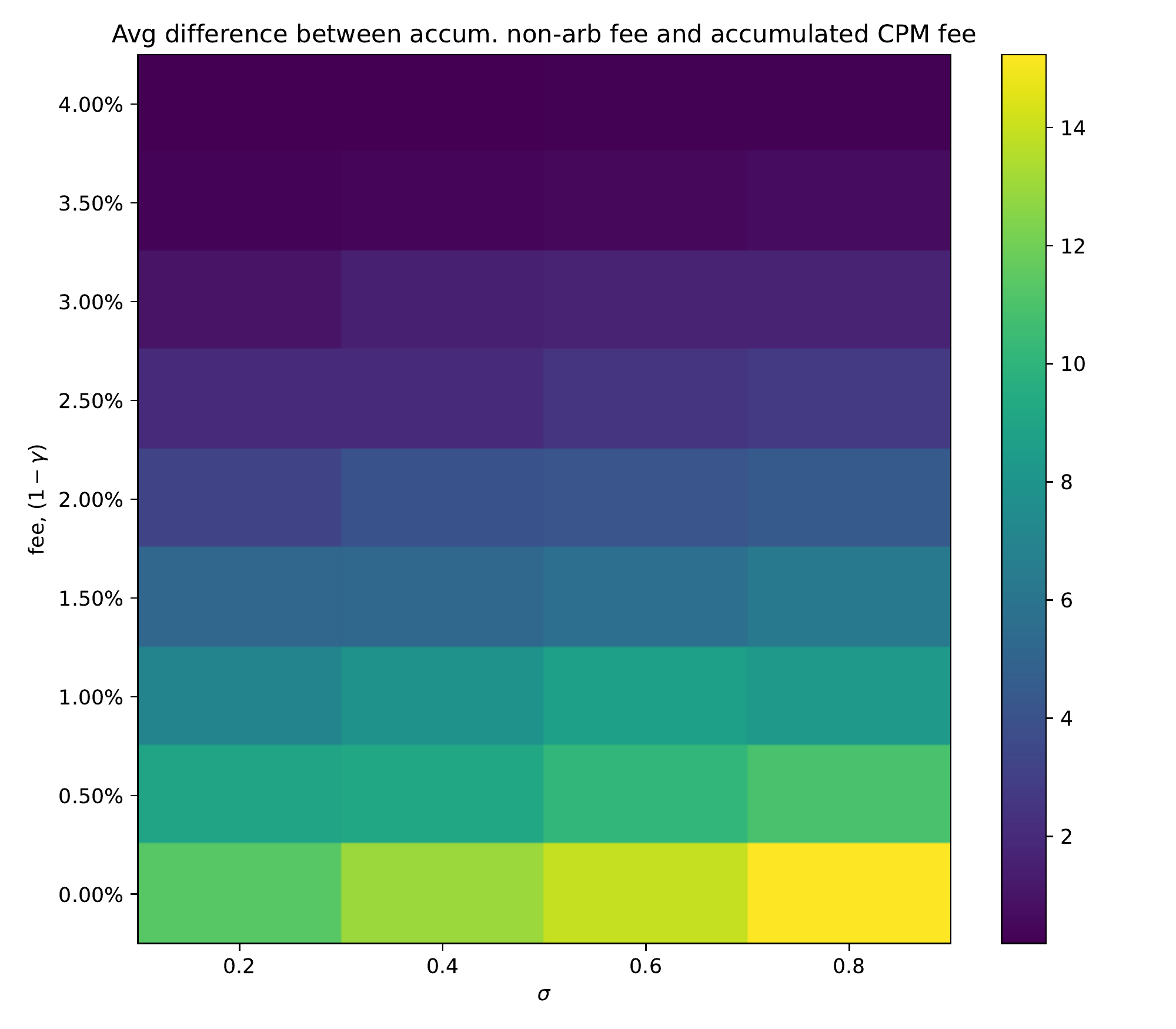}
         \caption{avg difference between non-arbitrage fees and CFM fees}
         \label{fig:avg diff fees}
     \end{subfigure}
     \hfill
     \begin{subfigure}[b]{0.45\textwidth}
         \centering
         \includegraphics[width=\textwidth]{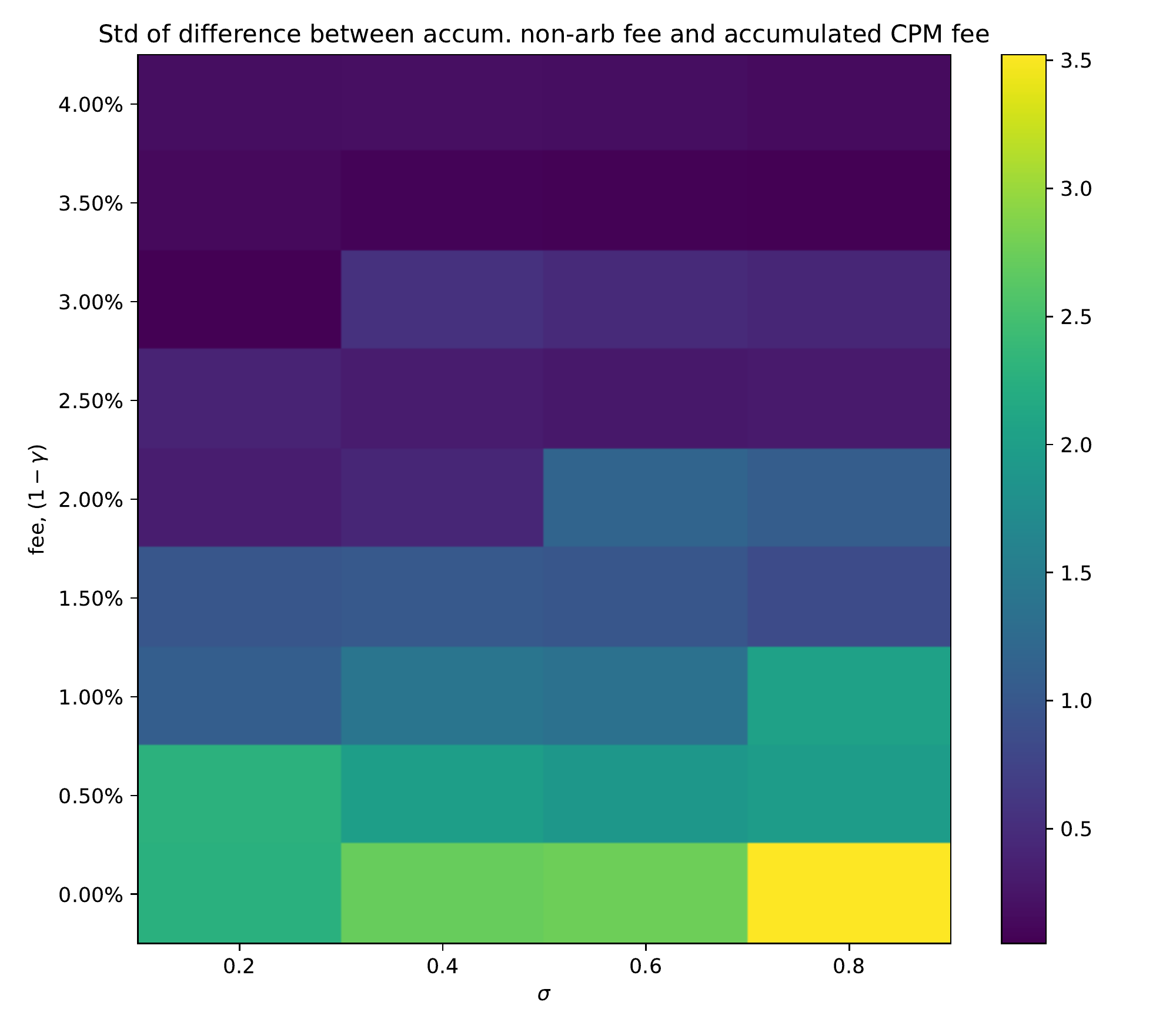}
         \caption{std of difference between non-arbitrage fees and CFM fees}
         \label{fig:std diff fees}
     \end{subfigure}
     \caption{Difference between premium fees and CFM fees for different values of $\gamma, \sigma$, positive values indicate fee income is not covering cost of hedging impermanent loss.}
     \label{fig:diff fees}
\end{figure}

\subsection{Literature review}
The CFM are gaining popularity in market design as they guarantee market liquidity (proportional to the amount of assets locked) and they have minimal storage and computational requirements, unlike e.g. limit-order-book-based markets.
This makes them ideal for environments where computational power and storage are at a premium like Ethereum~\cite{ethereum:2014}, or Cardano, Polkadot or Solana.

Running on blockchain gives advantages over centralised exchanges: transparency, pseudonymity, censorship resistance and security, see references in e.g.~\cite{loesch:2021}.
However blockchain-based CFMs suffer from a number of problems: namely questionable returns for LPs as has been already mentioned as well as traders (liquidity takers) being front-run by miners (known as miner-extractable-value)~\cite{daian:2019}, \cite{mevexplore:2021}.

Classical market clearing models suggest, see  Glosten \& Milgrom~\cite{glosten:1985}, Kyle~\cite{kyle:1985,kyle:1989} and Grossman \& Miller~\cite{grossman:1988}, that the volatility of the underlying asset plays an important role in how the market makers should set bid-ask spread (which is comparable to the pool fee as it imacts execution costs).
As we show in Theorem~\ref{thm critical rate} the fee income due to the CFM LPs is bounded above by a product of quadratic variation of the underlying asset (which is basically volatility) and convexity of the position being hedged. 

\subsection{Organisation of the paper}
Section~\ref{sec cfm thoery} includes all the theoretical findings of the paper.
Some readers may wish to skim through Section~\ref{sec cfm mechanics} which reviews CFMs and how liquidity provision works and fixes some notation used further in the paper.
Section~\ref{sec perp} recalls what a perpetual option is, Section~\ref{section arbitrage free model} proves the main theoretical result of the paper which is Theorem~\ref{thm critical rate}.
Section~\ref{sec abm} describes in detail the agent-based model used and discusses the findings.
Finally, in Section~\ref{sec conclusion}, we discuss the main findings, their limitations and possible further work.
A full implementation of the agent-based simulation is at\footnote{The link is deliberately broken to comply with the anonymous submission format.} \href{https://github.com/simtopia/cfm-sim}{\texttt{https://github.com/simtopia/cfm-sim}}.

\section{Liquidity provision in constant function markets}
\label{sec cfm thoery}

In this section, we will draw a connection between a constant function market (CFM) and a perpetual (American) options contract. 
We will particularly seek to derive the impermanent loss (also known as LVR) of a CFM, or equivalently the arbitrage-free streaming premium of the perpetual option.

\subsection{An overview of CFMs} 
\label{sec cfm mechanics}

A constant function market (CFM) is characterised by
\begin{enumerate}[i)]
\item The reserves $(x^1,x^2)\in \mathbb R^2_+$ describing amounts of assets in the pool. 
\item A trading function  $\Psi:\mathbb R_+^2 \rightarrow \mathbb R$ which determines the state of the pool after each trade according to the acceptable fund positions:
\begin{equation}
\left\{ (x^1,x^2) \in \mathbb R^2_+\, :\, \Psi(x^1,x^2) = \text{constant}\right\}.
\end{equation}
\item A trading fee $(1-\gamma)$, for $\gamma \in (0,1]$.
\end{enumerate}

For the purposes of this paper, we assume  $\Psi:\mathbb R_+^2 \rightarrow \mathbb R$ to be twice continuously differentiable and convex (see \cite{capponi2021adoption}, \cite{angeris2019analysis}). 

To acquire $\Delta x^1_t$ of asset $x^1$ at time $t$ a trader needs to deposit a quantity $\Delta x^2 = \Delta x^2(\Delta x^1)$ of asset $x^2$ into the pool, and pay a fee  $(1-\gamma) \Delta x^2$.\footnote{The fee in Uniswap-V3 is not added to the pool reserves \cite{adams2021uniswap}. This is in contrast to Uniswap-V2.}  $\Delta x^2$ and $\Delta x$ needs to satisfy the equation 
\begin{equation} \label{eq princing eqn}
	 \Psi(x^1_t - \Delta x^1, x^2_t + \Delta x^2) =
\Psi(x^1_t , x^2_t)\,.
\end{equation}
Once the trade is accepted the reserves are updated according to 
\begin{equation}
x^1_{t+1}= x^1_t - \Delta x^1 \quad \text{and} \quad x^2_{t+1}= x^2_t + \Delta x^2\,.
\end{equation}
The relative price of trading $\Delta x^1$ for  $\Delta x^2$ is defined as
\begin{equation}
    \frac{P_t^{1,CFM}(\Delta x^1)}{P_t^{2,CFM}(\Delta x^2)}:= \frac{\Delta x^2  }{ \Delta x^1}\, \quad \text{subject to }\quad  \Psi(x^1_t - \Delta x^1, x^2_t + \Delta x^2) =
\Psi(x^1_t , x^2_t)\,.
\end{equation}
Observe that 
\begin{equation}
\begin{split}
 0 =&  \Psi(x^1_t - \Delta x^1, x^2_t + \Delta x^2) - 
\Psi(x^1_t , x^2_t) \\
& \qquad =  - \partial_{x^1}\Psi(x^1_t,x^2_t) \Delta x^1 + \partial_{x^2}\Psi(x^1_t,x^2_t)\Delta x^2 + \mathcal O((\Delta x)^2)\,.
\end{split}
\end{equation}
Hence the relative price of trading an infinitesimal amount of $\Delta x^1$ for $\Delta x^2$ is given by 
\begin{equation}
 \frac{P_t^{1,CFM}}{P_t^{2,CFM}}:= \lim_{\Delta x^1\rightarrow 0} \frac{P_t^{1,CFM}(\Delta x^1)}{P_t^{2,CFM}(\Delta x^2)} 
 =  \frac{\partial_{x^1}\Psi(x^1_t,x^2_t)}{\partial_{x^2} \Psi(x^1_t,x^2_t)}  \,.
\end{equation}


Assume that there is an external market where assets $x^1$ and $x^2$ can be traded (without frictions) at the prices $S_t=(S_t^1,S_t^2)$. The no-arbitrage condition in the case of no fees ($\gamma=1)$ implies that
\begin{equation} \label{eq no arbitrage price}
  \frac{P_t^{1,CFM}}{P_t^{2,CFM}} = \frac{S_t^{1}}{S_t^{2}}
\end{equation}
 Conversely, if \ref{eq no arbitrage price} would not hold, then (in a market with no fees) there would be an arbitrage opportunity between CFM and the external market (assuming frictionless trading is possible), as it would be possible to purchase a combination of assets cheaply in one market, and then sell it in the other.

\begin{example}[GMM] \label{ex gmm}
Consider the trading function 
\begin{equation}
   \Psi(x^1,x^2)=(x^1)^{\theta}(x^2)^{1-\theta}
\end{equation}
for $\theta\in(0,1)$. The no arbitrage relationship  \eqref{eq no arbitrage price}, in GMM is given by
\begin{equation} \label{eq no arbitrage price gmm} 
  \frac{P_t^{1,CFM}}{P_t^{2,CFM}} =  \frac{S_t^{1}}{S_t^{2}}\,.
\end{equation}
The value of the liquidity pool at any time $t\in[0,\infty)$, is given by
\begin{equation}
\psi(S_t;x_t):= x^1_t \cdot S_t^1 + x^2_t \cdot S_t^2\,,   
\end{equation} 
and using \eqref{eq no arbitrage price gmm} we can show that 
\begin{equation}
   \psi(S_t;x_t) = \frac{1}{\theta} S_t^1 x^1_t\, \quad \text{or} \quad \psi(S_t)= \frac{1}{1-\theta} S_t^2 x^2_t\,. 
 \end{equation}
In Appendix~\ref{app gmm} we derive an alternative representation for $\psi_t$, that does not depend on $(x^1,x_t^2)$, 
\begin{equation}
    \psi(S_t) = \psi(S_0) \cdot \left(\frac{S_t^1 }{S_0^1} \right)^{\theta } \cdot  \left(  \frac{S_t^2 }{S_0^2} \right)^{1-\theta} \,. 
\end{equation}
\end{example}
The above example, also studied in \cite{evans2021liquidity}, demonstrates two key properties of CFMs:
\begin{itemize}
\item  A GMM automatically re-balances liquidity pools so that the value of the pools with asset  $x^1$  and $x^2$ is  $\theta \cdot \psi(S_t)$ and $(1-\theta)\psi(S_t)$, respectively
\item Providing liquidity to a CFM $\Psi$ is equivalent to entering a long position on a perpetual derivative on the underlying asset with the payoff dictated by the value (in terms of price) $\psi$. 
\end{itemize}

As observed in \cite{angeris2021replicating2}, \cite{angeris2021replicating}, for any CFM $\psi$ for any non-negative, non-decreasing, concave, 1-homogenous\footnote{That is,
 $\lambda \psi(S) =  \psi(\lambda S)$ for all $\lambda>0$ and $S$. Equivalently, we can assume that one of the assets in $S_t$ is the numeraire asset, in which case the 1-homogeneity assumption is not needed.} payoff function $\psi$, there exists a trading function $\Psi$, such that the value of the liquidity provision in CFM with with function $\Psi$ matches this payoff function. In other words, a CFM with appropriately designed trading function $\Psi$ dynamically adjusts portfolio held by liquidity providers so that the value of this portfolio, at any time, is described by the payoff function $\psi$. This gives us two ways to understand a CFM:
 \begin{itemize}
     \item Through the defining trading function $\Psi$, indicating valid combinations of the two assets.
     \item Through the valuation function $\psi$, indicating the value of the CFM pool in terms of the prices of the two assets.
 \end{itemize}

Concentrated liquidity, introduced in Uniswap V3, gives individual LPs control over what price ranges their capital is allocated to. This gives a practical way of creating liquidity provision that replicates a given function $\psi$, and has been observed and described in \cite{lambert2022panoptic}. 

\subsection{Connection with perpetual options}
\label{sec perp}
Given a payoff function $\psi$, there is a simple connection between liquidity provision in a CFM and investment in a perpetual American option.

We begin with a precise definition of a perpetual American contract with streaming premium. As we shall see, liquidity provision in a CFM is equivalent to entering a long perpetual derivative with the payoff dictated by the function $\psi$. This derivative is written on the underlying assets $S=(S_t)_{t\geq 0}$ traded in the pools.  We follow the definition from \cite{angeris2022primer}.

\begin{definition}
A perpetual contract with streaming premium, written on assets with price vector $S$, with payoff function $\psi:\mathbb R^n \to \mathbb R$, is a agreement between two parties, referred to as the long side and short side. The long side has the right to terminate the contract at any time $t \geq 0$, at which point it will receive a payment of $\psi(S_t)$. In return, the long-side must pay to the short side $\psi(S_0)$ at the time $t = 0$ of inception as well as a continuous cash-flow of $ (g_t)_{t\geq 0}$ per unit time, referred to as the streaming premium, up until the contract is terminated.
\end{definition}

In order to highlight the connections between a CFM and a perpetual American option, we make the following observations. In a CFM,  a liquidity provider (CFM-LP)
\begin{itemize}
    \item  initially deposits assets with value $\psi(S_0)$,
    \item receives fees at the rate $f_t$ per unit time (which may vary). These fees may be withdrawn (under the Uniswap v3 protocol) at any time,
    \item at some future time $\tau$ (of the CFM-LPs choosing), may withdraw their assets, which will have value $\psi(S_\tau)$.
\end{itemize}
In a perpetual American option with streaming premium and payoff $\psi$, an investor purchasing the option
\begin{itemize}
    \item initially purchases the option, for a cost $\psi(S_0)$,
    \item receives the streaming premium at a rate $g_t$ per unit time (which may vary). This streaming premium may be positive or negative, and is received immediately,
    \item at some future time $\tau$ (of the investor's choosing), may exercise the option, which rewards them with value $\psi(S_\tau)$.
\end{itemize}
As we can see, assuming fees in the CFM are instantaneously predictable (i.e. the fee to be received from the CFM can be accurately estimated in advance), a no arbitrage argument suggests that we must have $f_t=g_t$, as otherwise one asset is strictly better than the other in every state of the world (over some short time horizon), which implies that an efficient market will focus all its trading in the better of these alternatives.

\subsection{Model and assumptions}
\label{section arbitrage free model}

We make the following assumptions about the market:
\begin{itemize}
\item The agent can borrow and lend any amount of cash at the riskless rate.
\item The agent can buy and sell any amount of assets $x^1$ and $x^2$. 
\item The above transactions do not incur any transaction costs and the size of a trade does not impact the prices of the traded assets. 
\end{itemize}

We denote the risk free rate (which may be stochastic) by $(r_t)_{t\geq 0}$
and model the money market as
\begin{equation}
\label{eq bond}
    dB_t = r_t B_t dt\,, \qquad B_0\geq 0\,. 
\end{equation}
We assume $r_t\ge 0$. We further denote the drift and diffusion coefficients (again, possibly stochastic) of the risky asset shadow prices by $\mu=(\mu_t^1,\mu_t^2) \in \mathbb R^2$ and \[
\sigma=(\sigma_t^{(1,1)}, \sigma_t^{(1,2)},\sigma_t^{(2,1)},\sigma_t^{(2,2)}) \in \mathbb R^4_+
\] 
respectively, and model the shadow price of the risky asset $(S^i_t)_{t\geq 0}$ by
\begin{equation} \label{eq risky asset}
    dS_t^{i}= \mu^i_t S_t^i dt + \sum_{j=1}^d \sigma^{(i,j)}_t S_t^i dW_t, \quad S_o^i\geq 0\,,
\end{equation}
where $W=(W_t^1,W_t^2)$ is a $2$-dimensional Brownian motion. We assume that $(\mu,\sigma)$ are sufficiently regular that a (unique strong) solution to \eqref{eq risky asset} exists.

\begin{remark}
By allowing the drift and diffusion coefficients to be arbitrary processes we are just saying that the prices are non-negative and continuous: our framework incorporates a rich family of common models such as Black--Scholes, Heston, SABR or local stochastic volatility models with possibly path dependent coefficients.
\end{remark}

We will proceed to derive an arbitrage-free fee bound in a similar way to the construction of the predictable loss in~\cite{milionis2022automated}, and to classical arguments for no-arbitrage pricing in financial markets.

\begin{assumption} \label{as psi}
The value of the pool as a function of external price process, $s \mapsto \psi(s)$ is twice continuously differentiable. 
\end{assumption}
Assumption \ref{as psi} holds in the case when the fee $(1-\gamma)=0$ and arbitrageurs  continuously close the gap between $S_t$ and the price of the CFM, as demonstrated in \cite{milionis2022automated}. Hence we expect the the arbitrage-free fee derived below to be a good proxy for CFMs that charge small fee (e.g Uniswap). If the CFM charges a fee $(1-\gamma)$ which is significantly larger than $0$ then the value of the pool is a function of the reserves and the price one would use for the conversion (external price $S_t$ or pool-implied price $P^{2,CFM}_t$), see Remark~\ref{rmk path depdend pool value} for more details. 

Consider the wealth process $Z$ of an agent who
\begin{itemize}
    \item begins with zero capital\footnote{This is simply to avoid having to account for the interest they should earn on their initial capital.},
    \item initially borrows a quantity $\psi(S_0)$ at the risk-free interest rate, which they use to establish a CFM position (which they can do by trading in the liquid market to obtain the desired risky assets for the pool),
    \item until time $t$, trades in the liquid market hold a quantity $(\Delta_s^i)$ of each asset at time $s$.
\end{itemize}
The dynamics of $Z_t$ are given by 
\[
Z_t = \underbrace{\psi(S_t)-\psi(S_0)}_{\text{CFM gains}}+\underbrace{\int_0^t\sum_{i=1}^2 \Delta^i_s dS^i_s}_{\text{trading gains}} + \underbrace{\int_0^t\Big(Z_s - \sum_{i=1}^2 \Delta^i_s S^i_s\Big)r_s ds}_{\text{interest payments on uninvested wealth}} + \underbrace{\int_0^t f_s ds}_{\text{CFM fees}}.
\]
We assumed $\psi$ is smooth, hence we can apply It\^o's lemma to obtain
\[\begin{split}
  Z_t &= \sum_{i=1}^2 \int_0^t\Big[\partial_i \psi(S_s) + \Delta^i_s\Big]dS_s^i
 + \frac{1}{2}\sum_{i,j=1}^2 \int_0^t \partial_i \partial_j \psi(S_s)d\langle S^i,S^j\rangle_s + \int_0^t\Big[\Big(Z_s - \sum_{i=1}^2 \Delta^i_s S^i_s\Big)r_s+f_s\Big] ds.
\end{split}
\]
By setting $\Delta_s^i = -\partial_i \psi(S_s)$, we eliminate the first term, giving
\[\begin{split}
  Z_t &= \frac{1}{2}\sum_{i,j=1}^2 \int_0^t \partial_i \partial_j \psi(S_s)d\langle S^i,S^j\rangle_s + \int_0^t\Big[\Big(Z_s + \sum_{i=1}^2 (\partial_i \psi(S_s) ) S^i_s\Big)r_s+f_s\Big] ds.
\end{split}
\]
As the quadratic variation is  $d\langle S^i,S^j\rangle_s = (\sigma_s\sigma_s^\top)^{(i,j)} S_s^iS_s^jds $, this simplifies to
\[
  \frac{dZ_s}{ds} = \frac{1}{2}\sum_{i,j=1}^2 \partial_i \partial_j \psi(S_s)(\sigma_s\sigma_s^\top)^{(i,j)} S_s^iS_s^j + \Big(Z_s + \sum_{i=1}^2 (\partial_i \psi(S_s) ) S^i_s\Big)r_s+f_s.
\]
If this quantity is positive, then we have a strategy which earns money with probability one over a short time period, that is, an arbitrage. Therefore, after rearrangement, we know that 
\[
  f_s  + Z_sr_s\le -\frac{1}{2}\sum_{i,j=1}^2 \partial_i \partial_j \psi(S_s)(\sigma_s\sigma_s^\top)^{(i,j)} S_s^iS_s^j - \Big(\sum_{i=1}^2 (\partial_i \psi(S_s) ) S^i_s\Big)r_s.
\]
As $Z$ is increasing in $f$ (an agent who earns more fees will be wealthier) and we assume $r\ge 0$, there is a unique value of $f$ such that above equation is an equality. 
With this CFM trading fee income per unit of time, we have $dZ_s/ds = 0$ and $Z_0=0$,  and hence $Z_s = 0$. 
We denote this critical value\footnote{\cite{faycal} give a similar calculation, to obtain a closely related quantity, which they call the permanent loss of the CFM.}
\[
\hat f_s = -\frac{1}{2}\sum_{i,j=1}^2 \partial_i \partial_j \psi(S_s)(\sigma_s\sigma_s^\top)^{(i,j)} S_s^iS_s^j - \Big(\sum_{i=1}^2 (\partial_i \psi(S_s) ) S^i_s\Big)r_s.
\]
We have thus shown the following.
\begin{theorem}
\label{thm critical rate}
Given the model of the market given by~\eqref{eq bond} and~\eqref{eq risky asset} and with the assumptions stated in Section~\ref{section arbitrage free model} we know that CFM fee income rate $f_t$ (which depends on the trade flow and the fee $1-\gamma$) must satisfy $f_t \leq \hat f_t$ where this theoretical upper bound on CFM fee income rate is
\begin{equation}
\label{eq fee rate}
\begin{split}
\hat f_t = & -\frac{1}{2}\sum_{i,j=1}^2  \partial_i \partial_j \psi(S_t) \frac{d\langle S^i,S^j\rangle_t}{dt}-\Big(  \sum_{i=1}^2\partial_i \psi(S_s) S_t^i\Big)  \frac{d \log B_t}{dt}\,.
\end{split}
\end{equation}
\end{theorem}

Notice that the critical CFM fee income rate depends on the trading function via $\partial_i \partial_j \psi(S)$ (which is negative, as $\psi$ is concave as long as the trading function $\Psi$ is convex) and the quadratic variation of traded asset (which is implicitly related to the level of trading activity). 
On the other hand the actual CFM fee income rate depends on the trade flow in the CFM and the fee the pool sets i.e. $1-\gamma$.
It is not clear that one could bring $f_t$ in line with $\hat f_t$ as increasing $1-\gamma$, the fee charged by the CFM will likely lead to decreased trade flow.

Note that this argument only uses a long-position in the CFM, and only establishes\footnote{The presentation above assumes the agent starts at time $0$, and derives the critical fee rate on this basis. 
To obtain the inequality bound $f_t\leq \hat f_t$ for all times, we formally have to consider starting with zero capital at a time where the inequality is not satisfied, and showing that this gives a short-term arbitrage opportunity.} the inequality $f_t\leq \hat f_t$. 
If it were possible to perfectly short-sell the CFM (or the corresponding perpetual option), then a similar argument would yield the converse inequality. This may explain why the fee rate in Uniswap v3, is systematically below the critical fee rate, which is related to liquidity provision in CFMs yielding persistently poor returns.


\begin{example}
\label{example arbitrage}
\label{ex gmm cont}
 For clarity of presentation we set the risk free interest rate $r=0$,  and assume that asset $S^2$ is a numeraire (hence the agent only invests in asset $S=S^1$). This means that
\begin{equation}
\psi(S_t) = \psi(S_0) \left(\frac{S_t }{S_0} \right)^{\theta } \,. 
\end{equation}
The second derivative of $\psi$ is given by $\partial^2_S \psi(S_t) = \theta (\theta -1)\frac{\psi(S_0)}{S_0^2} \left( \frac{S_t}{S_0} \right)^{\theta -2} $, which is negative since $\theta\in(0,1)$. The critical fee rate is then given by 
\begin{equation}
    \begin{split}
\hat f_t &= \frac{-1}{2}  \theta (\theta -1)\frac{\psi(S_0)}{S_0^2} \left( \frac{S_t}{S_0} \right)^{\theta -2} \sigma_t^2S_t^2\\
&= \frac{\theta(1-\theta)}{2}\sigma_t^2 \psi(S_0)\left(\frac{S_t}{S_0}\right)^{\theta}   =\frac{\theta(1-\theta)}{2}\sigma_t^2\psi(S_t)\geq 0\,.
\end{split}
\end{equation}

\end{example}

As mentioned above, by analyzing the data in the Uniswap v3 pool (e.g. \cite{faycal}) one can see that, in general, the fee rate is below critical fee rate. Nevertheless, since one cannot simply short position at the Uniswap (unless using an over-the-counter bespoke arrangement) it is not clear how one could realise a potential arbitrage opportunity when $f < \hat f$.

\begin{remark}[Comments about the derivation of the arbitrage-free fee rate bound] 
\label{rmk path depdend pool value}

The derivation of the arbitrage-free fee rate bound is done under the Assumption \ref{as psi} which, as we explained, holds in the case when $(1-\gamma)= 0$.  

If for each trade the CFM charges a fee it can be easily shown that this optimal arbitrage trades will not lead to condition~\eqref{eq no arbitrage price}. We demonstrate this on Figure \ref{fig:price gamma 0.96}. Indeed the arbitrageur who trades $(\Delta x^1,\Delta x^2)$ when reserves are $(x^1,x^2)$ will incur cost $f(\gamma,x^1,x^2,\Delta x^1, \Delta x^2)$  and  solves the optimisation problem  $\Delta x^{2, *} := \max (\Delta x^2_{\text{CFM, CEX}}, \Delta x^2_{\text{CEX, CFM}})$, where
\begin{equation}\label{eq:optimisation1 arb}
\begin{split}
\Delta x^2_{\text{CFM, CEX}} := &  \argmax_{\Delta x^2} \Delta x^1 \cdot S^1 - \Delta x^2 - f(\gamma,,x^1,x^2,\Delta x^1, \Delta x^2)   \\
\text{such that } & \Psi(x_1 - \Delta x^1, x_2 + \Delta x^2) =\Psi(x_1 , x_2 ), \quad \Delta x^2 \geq 0, \\
 & \Delta x^1 \cdot S^1 - \Delta x^2 - f(\gamma,x^1,x^2,\Delta x^1, \Delta x^2) \geq 0
\end{split}
\end{equation}
and 
\begin{equation}\label{eq:optimisation2 arb}
\begin{split}
\Delta x^2_{\text{CEX, CFM}} := & \argmax_{\Delta x^2} \Delta x^2 - \Delta x^1 S^1 - f(\gamma,x^1,x^2,\Delta x^1,\Delta x^2)   \\
\text{such that } & \Psi(x_1 + \Delta x^1, x_2 - \Delta x^2) =\Psi(x_1 , x_2 ), \quad \Delta x^2 \geq 0, \\
 &  \Delta x^2 - \Delta x^1 S^1 - f(\gamma,x^1,x^2,\Delta x^1,\Delta x^2) \geq 0
\end{split}
\end{equation}
where the subindexes of $\Delta x^2_{\cdot, \cdot}$ denote the order of the markets where the arbitrageur goes short and long in asset $x^2$. If $\Delta x^{2,*} = 0$, then there is no possible trade for which an arbitrageur makes a profit.
When condition~\eqref{eq no arbitrage price} does not hold, the model to calculate the arbitrage fee rate bound $\hat f_t$  needs to be generalized. A first step in this direction is to re-rewrite the value function as a function of $(x^1,x^2,S_t)$. Furthermore, using Implicit Function Theorem for trading function $\Psi(x^1,x^2)$, there is function $g$ such that $x^2 = g(x^1)$. Hence, assuming $x^1$ is the numeraire so that $S_t^1=1$,  
\[
\psi(x^1_t,S_t) = x^1_t  + g(x^1_t) \cdot S^2_t\,.
\]
Applying It\^o formula to $\psi$ we get
\[
\begin{split}
d\psi(x_t^1, S^2_t) = & \partial_x \psi(x_t^1, S^2_t) dx_t^1 + \partial_S \psi(x_t^1, S^2_t) dS^2_t + \partial_x \partial_S \psi(x_t^1, S^2_t) d\langle x^1, S^2\rangle_t \\
& + \frac1{2}\partial_x^2 \psi(x_t^1, S^2_t) d\langle x^1, x^1\rangle_t  + \frac1{2}\partial_S^2 \psi(x_t^1, S^2_t) d\langle S^2, S^2\rangle_t\,.
\end{split}
\]
One can see that to analyse the evolution of the value of the pool one needs to model the evolution of reserves and its dependence on $S$. Instead of performing analysis under various sets of assumptions we turn to simulations.  
\end{remark}

\section{An Agent-based model}
\label{sec abm}

We design an Agent-Based-Model with the aim to compare the accumulated fees by a CFM defined as in Example \ref{ex gmm} with $\theta=1/2$ in a period of time $[0,T]$ against the accumulated non-arbitrage fee derived in Section \ref{section arbitrage free model}. 

\subsection{Model description}

We consider the CFM with $x^1$ being the numeraire. Price discovery occurs in a second reference market, where $S_t$ denotes the price of asset $x^2$ in terms of the numeraire $x^1$ at time $t$.  We model the price $S_t$ by a Geometric Brownian Motion
\[
dS_t = \sigma S_t dW_t, \quad S_0>0.
\]
Whenever the price of asset $x^2$ in terms of the numeraire $x^1$ in the CFM is different than $S_t$, there might be an arbitrage opportunity. An arbitrageur will automatically make the necessary trades to maximise their profit. Furthermore we model the arrival of liquidity takers by Poisson process $(N_t)_t$ with intensity $\lambda$. We run the simulation of discrete times $\Pi := \{0,t_1,t_2,...,T\}$ and initialise the simulation with reserves $(x^1_0,x^2_0)$ and fee $(1-\gamma)$; initial  price $S_0$ and volatility $\sigma$; liquidity takers arrival intensity $\lambda$. We run the following simulation. 

    For every $t\in\Pi$:
    \begin{itemize}
        \item The arbitrageur makes the necessary trades resulting from optimisations \eqref{eq:optimisation1 arb} \eqref{eq:optimisation2 arb} in the CFM and in the reference market to make a risk-free profit.
        \item If $N_t - N_{t-} = 1$ then a price-sensitive liquidity trader arrives and trades $\Delta_t\in\mathbb R$ sampled from $\mathcal N(0,1)$ of asset $x^1$. The noise trader chooses the market with better execution price with some high probability $p$, i.e. we allow for the noise trader to be irrational with probability $(1-p)$\footnote{Perfectly rational liquidity trader would consider splitting the order to optimise the execution cost. We omit this extension in our simulations.}.
    \end{itemize}

We take $\sigma \in \{0.2,0.4,0.6,0.8\}, \lambda \in \{50,75,100\}$ and fix $S_0 = 10, x_0^1 = 100, x_0^2=10$ and run the above simulation 100 times for different values of $\gamma \in [0.96,1]$.

\subsection{High fee and low fee $1-\gamma$} 

\begin{figure}[ht]
 \centering
 \includegraphics[width=\textwidth]{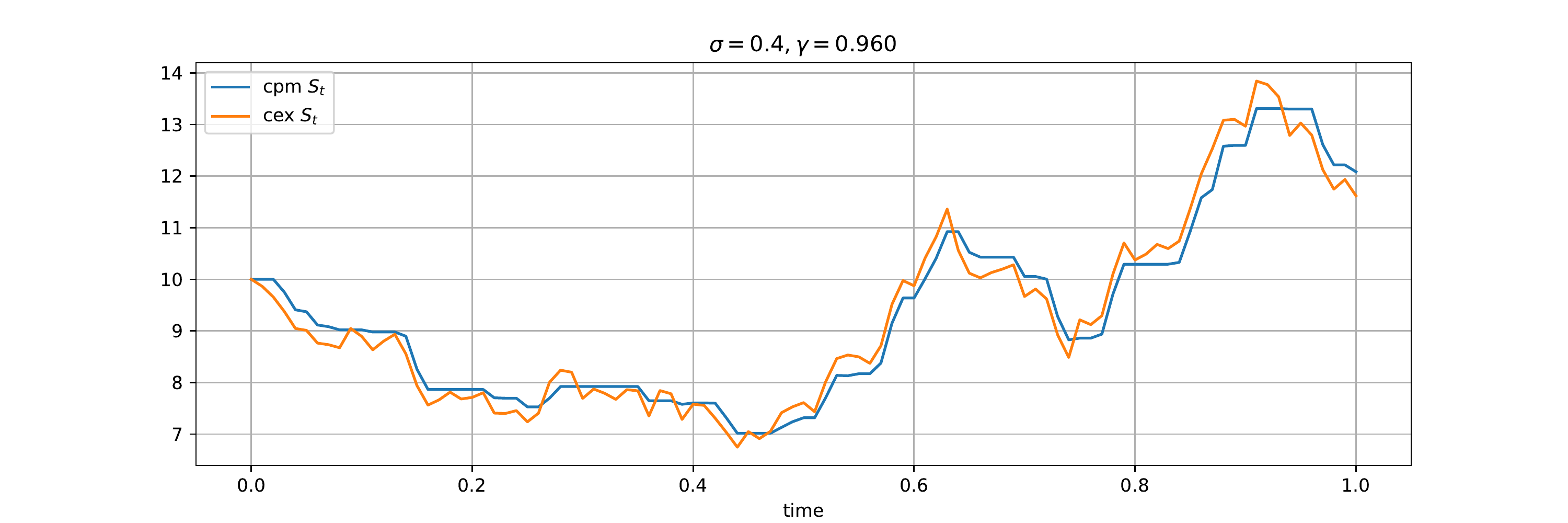}
 \caption{Price processes for a simulation with $\gamma = 0.96, \sigma=0.4$}
 \label{fig:price gamma 0.96}
\end{figure}

\begin{figure}[ht]
 \centering
 \includegraphics[width=\textwidth]{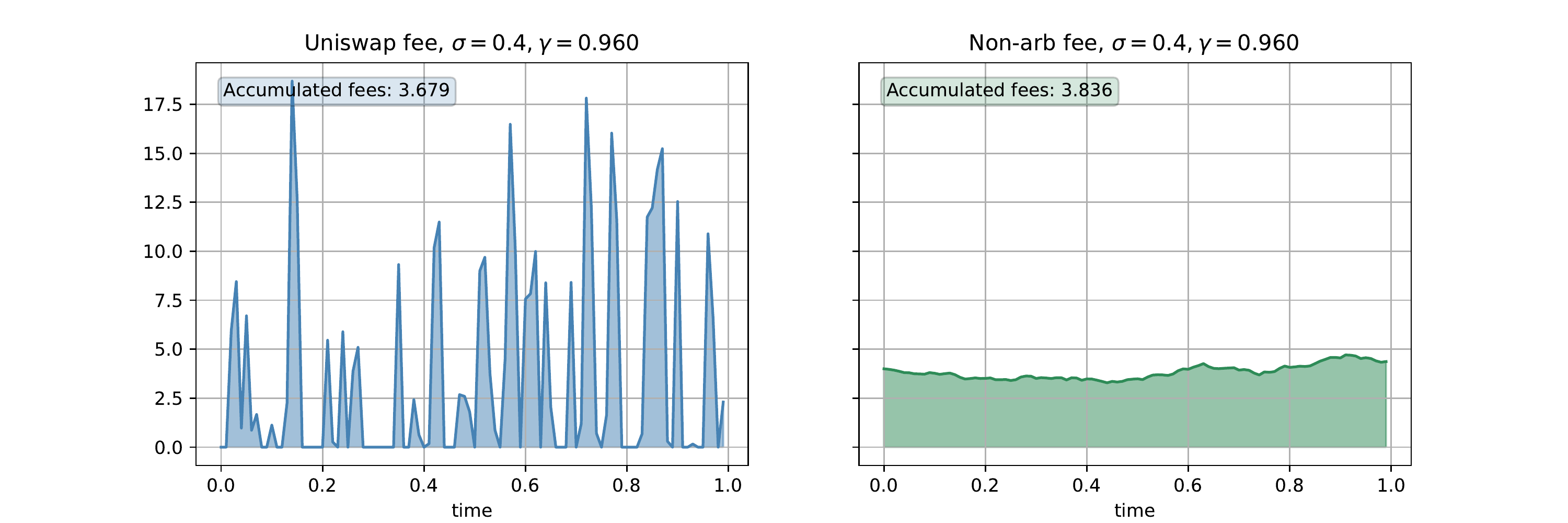}
 \caption{CFM fees (left) and non-arbitrage fees (right) for a simulation with $\gamma = 0.96, \sigma=0.4$}
 \label{fig:fees gamma 0.96}
\end{figure}

\begin{figure}[ht]
     \centering
     \includegraphics[width=\textwidth]{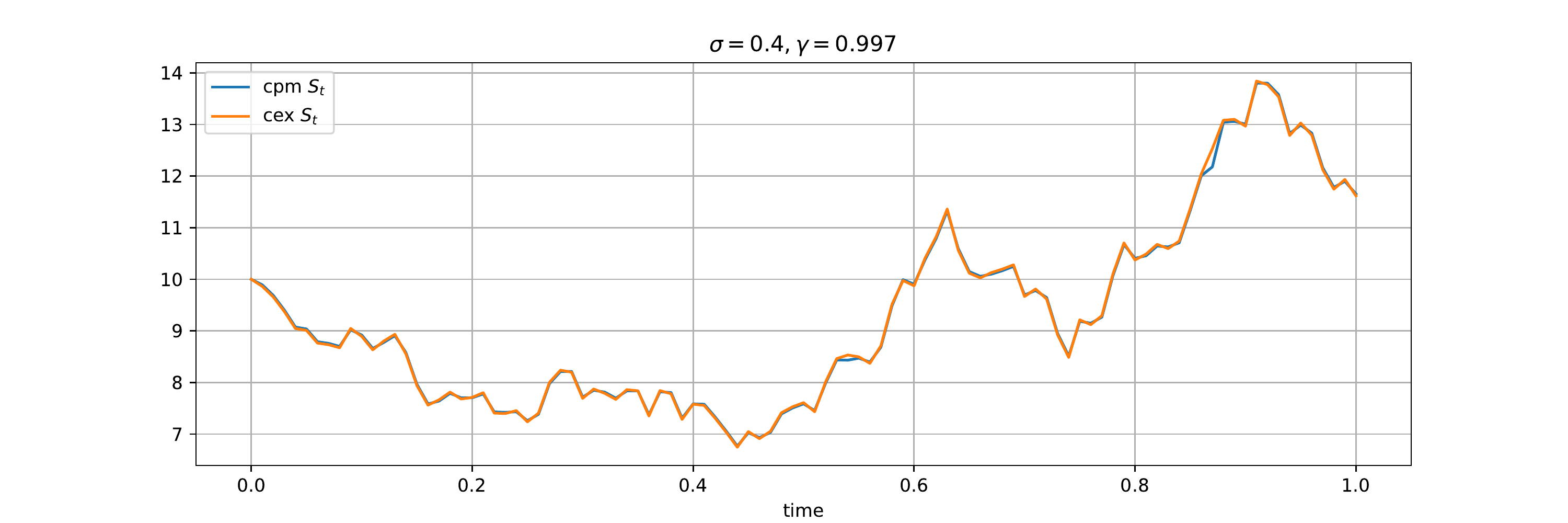}
     \caption{Price processes for a simulation with $\gamma = 0.997, \sigma=0.4$}
     \label{fig:price gamma 0.997}
 \end{figure}

\begin{figure}[ht]
     \centering
     \includegraphics[width=\textwidth]{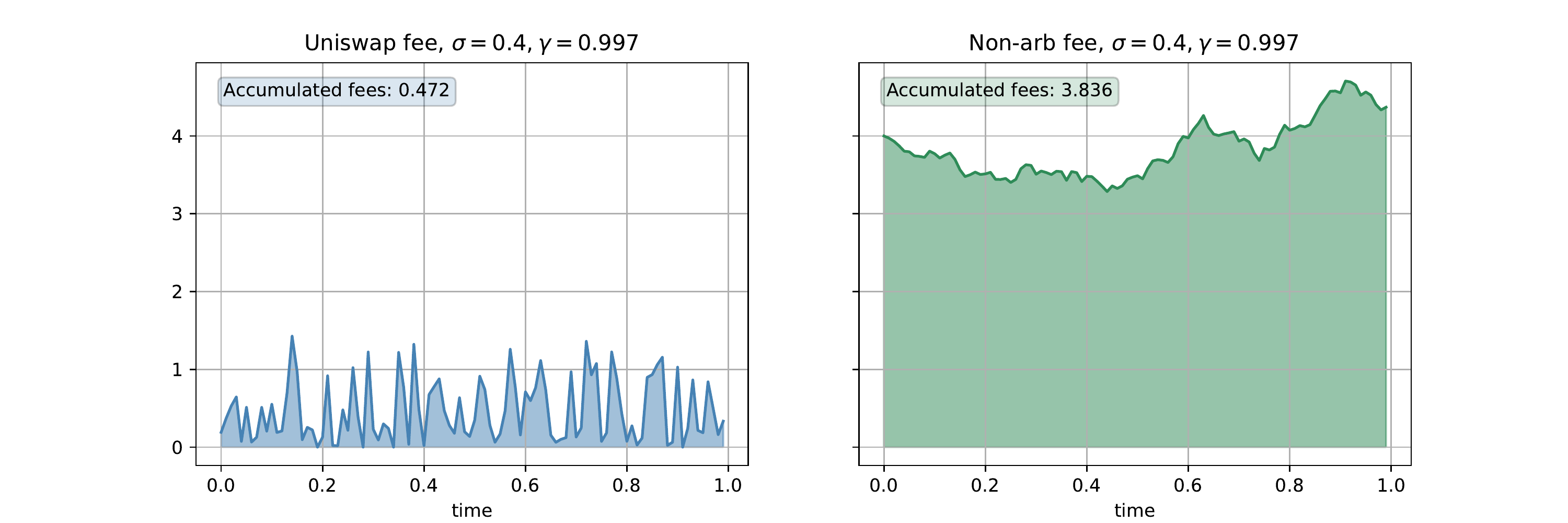}
     \caption{CFM fees (left) and non-arbitrage fees (right) for a simulation with $\gamma = 0.997, \sigma=0.4$}
     \label{fig:fees gamma 0.997}
 \end{figure}

Figures \ref{fig:price gamma 0.96}, \ref{fig:fees gamma 0.96}, \ref{fig:price gamma 0.997} and \ref{fig:fees gamma 0.997} consider two simulated scenarios:  high fee value $\gamma = 0.96$ ( i.e  fee $4\%$);  low fee value $\gamma=0.997$ of ( i.e fee $0.3\%$), $\lambda=50$ and $\sigma=0.4$. Figures \ref{fig:price gamma 0.96} and \ref{fig:price gamma 0.997} provide the evolution of the price process for the two simulated scenarios in the CFM for the same price process $S_t$ in the reference market. The smaller the fee (the higher the value of $\gamma$) the closer the CFM price process depicted in blue follows the reference market price $S_t$. For these two scenarios, the CFM and arbitrage-free fee rates are shown in Figures \ref{fig:fees gamma 0.96} and \ref{fig:fees gamma 0.997}, and the area under the fee rates corresponds to accumulated fees in that particular simulation. High values of fees (low values of $\gamma$) indicate less arbitrage opportunities, and therefore the number of trades will be lower than in the low fees scenarios. Hence increasing the fee rate does not necessarily lead to higher fee income. However, in higher fees regimes, the trade sizes will necessarily be higher (there will only be an arbitrage opportunity when the gap between the CFM price and the reference price is big enough to compensate for the fee). 

Fixing $\lambda=50$, Figure \ref{fig:diff fees} aggregates the generated results for all the considered $\gamma\in[0.96,1], \sigma\in\{0.2,0.4,0.6,08\}$. Figure~\ref{fig:avg diff fees}  indicates that higher trades will bring higher fees on average, yet they will not be able to cover the cost of hedging the impermanent loss, which increases with the volatility $\sigma$.

 \hfill


\subsection{Impact of noise trader arrival rate $\lambda$}

\begin{figure}[h]
    \centering
    \includegraphics[width=\textwidth]{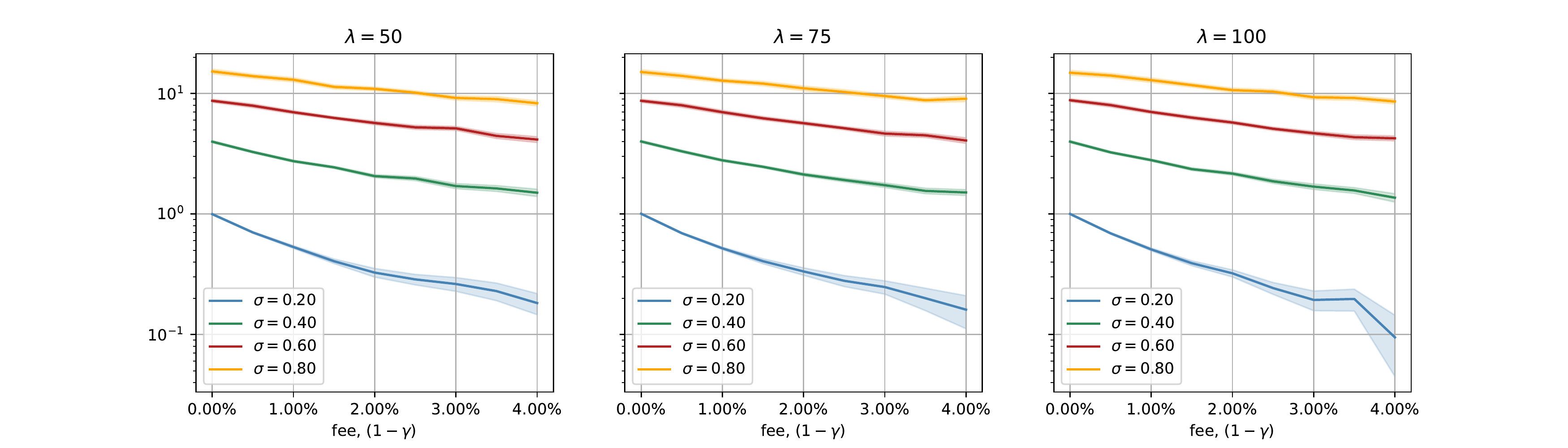}
    \caption{Difference between non-arbitrage fees and CFM fees for different values of $\gamma, \sigma, \lambda$}
    \label{fig:agg results}
\end{figure}

 Figure \ref{fig:agg results} aggregates all the simulations for three considered intensity regimes for the noise trader. Again, in average the CFM fee does not cover the cost of hedging impermanent loss and, furthermore, this is not impacted by the arrival intensity of noise traders.

\section{Conclusion and future work}
\label{sec conclusion}
In Theorem~\ref{thm critical rate} we have derived an upper bound on the fee income for an LP placing liquidity in a CFM which does not lead to arbitrage when compared to the cost of hedging the risk of the underlying price moves (referred to as impermanent loss).
These findings are augmented by agent based simulations which indicate that at least in some market regimes LP providers in CFMs are not being adequately compensated for the impermanent loss risk they take on. 

The theoretical model used to derive Theorem~\ref{thm critical rate} makes a number of  simplifying assumptions that are common in financial mathematics but on top of that it only applies in the low fee regime of CFMs, as discussed in Remark~\ref{rmk path depdend pool value}. 
To better study the higher fee regimes one would need to jointly model the trade flow and its impact on the underlying price or to model the joint evolution of the reserves and the underlying asset price. 
This, mathematically more challenging problem, is left for future work. 

The agent-based simulation could be exctended in a number of ways. 
More realistic agent order flow could be based directly on empirical data e.g. from~\cite{miori:2022}. 
The liquidity takers could not only choose the venue which provides better execution price but could split their orders for lowest overall execution cost.
Furthermore, most arbitrageurs do not to realise the arbitrage in one ``step''; instead they should consider multi-step optimisation with penalty for inventory leading to dynamic-programming problems (or, if we assume that the model of underlying price is unknown, to a reinforcement learning problem).
Again, this more involved investigation is left for future work.

\appendix 
\section{Derivation of $\psi$ for the Geometric Mean Market} \label{app gmm}

Consider the trading function 
\begin{equation}
   \Psi(x^1,x^2)=(x^1)^{\theta}(x^2)^{1-\theta}
\end{equation}
for $\theta\in(0,1)$. In the setting with no fees, $\gamma=1$, we have
$(x^1_t)^{\theta}(x^2_t)^{1-\theta}=(x^1_0)^{\theta}(x^2_0)^{1-\theta} $.
The no arbitrage relationship  \eqref{eq no arbitrage price}, in GMM is given by
\begin{equation} \label{eq no arbitrage price gmm} 
  \frac{P_t^{1,CFM}}{P_t^{2,CFM}} =\frac{\theta x^2_t}{(1-\theta)x^1_t} =  \frac{S_t^{1}}{S_t^{2}}\,.
\end{equation}
The value of the liquidity pool at any time $t\in[0,\infty)$, is given by
\begin{equation}
\psi(S_t):= x^1_t \cdot S_t^1 + x^2_t \cdot S_t^2\,.  
\end{equation}
Note that, under no arbitrage and no fee assumptions, it makes no difference whether the accounting is being done in $S$ or $P$. 

Using \eqref{eq no arbitrage price gmm} we can show that 
\begin{equation}
   \psi(S_t) = \left( \frac{1-\theta}{\theta} +1 \right) S_t^1 x^1_t = \frac{1}{\theta} S_t^1 x^1_t\,
\end{equation}
or equivalently 
\begin{equation}
   \psi(S_t) = \left( \frac{\theta}{1-\theta} +1 \right) S_t^2 x^2_t = \frac{1}{1-\theta} S_t^2 x^2_t\,. 
\end{equation}
From here we see that the value of the sub-pools with assets  $x^1$ and $x^2$ are  $\theta \cdot \psi(S_t)$ and $(1-\theta)\psi(S_t)$, respectively. 

Next, we derive an alternative representation for $V_t$ that does not depend on $(x^1,x_t^2)$. 
To do that, note that 
\begin{equation}
1 = \frac{\psi(S_t)}{\psi(S_t)} = \frac{S_t^2 x^2_t}{1-\theta} \frac{\theta}{S_t^1 x^1_t}.
\end{equation}
Hence
\begin{equation}
    \begin{split}
       \psi(S_t) & = \left(\frac{S_t^1 x^1_t}{\theta} \right)^{\theta } \cdot \left(\frac{S_t^1 x^1_t}{\theta} \right)^{1-\theta } \\
       &= \left(\frac{S_t^1 x^1_t}{\theta} \right)^{\theta } \cdot \left(\frac{S_t^1 x^1_t}{\theta} \right)^{1-\theta }\cdot
       \left(  \frac{S_t^2 x^2_t}{1-\theta} \frac{\theta}{S_t^1 x^1_t}\right)^{1-\theta} \\
       &= (x^1_t)^\theta (x^2_t)^{1-\theta} \left(\frac{S_t^1 }{\theta} \right)^{\theta } \cdot  \left(  \frac{S_t^2 }{1-\theta} \right)^{1-\theta}.
    \end{split}
\end{equation}
Since 
\begin{equation}
(x^1_t)^\theta (x^2_t)^{1-\theta} = (x^1_0)^\theta (x^2_0)^{1-\theta}\,, \quad \text{and} \quad
\psi(S_0)= (x^1_0)^\theta (x^2_0)^{1-\theta} \left(\frac{S_0^1 }{\theta} \right)^{\theta } \cdot  \left(  \frac{S_0^2 }{1-\theta} \right)^{1-\theta} \,,
\end{equation}
we have
\begin{equation}
    \psi(S_t) = \psi(S_0) \cdot \left(\frac{S_t^1 }{S_0^1} \right)^{\theta } \cdot  \left(  \frac{S_t^2 }{S_0^2} \right)^{1-\theta} \,. 
\end{equation}

\bibliographystyle{abbrv}
\bibliography{bibliography.bib}

\end{document}